\documentclass[%
 reprint,
 amsmath,amssymb, aps,
]{revtex4-1}

\usepackage{graphicx}
\usepackage{dcolumn}
\usepackage{bm}
\usepackage[mathlines]{lineno}

\begin{document}

\preprint{APS/123-QED}

\title{Quantitative analysis of pedestrian counterflow in a cellular automaton
model}

\author{Stefan Nowak} \email{sn@thp.uni-koeln.de} \author{Andreas
  Schadschneider}%
\email{as@thp.uni-koeln.de} \affiliation{%
  Institute for Theoretical Physics, Universit\"{a}t zu K\"oln,
  D-50937 K\"oln, Germany }%

\date{\today}

\begin{abstract}
  Pedestrian dynamics exhibits various collective phenomena. Here we
  study bidirectional pedestrian flow in a floor field cellular
  automaton model. Under certain conditions, lane formation is
  observed.  Although it has often been studied qualitatively, e.g.\ 
  as a test for the realism of a model, there are almost no
  quantitative results, neither empirically nor theoretically. As
  basis for a quantitative analysis we introduce an order parameter
  which is adopted from the analysis of colloidal suspensions.  This
  allows to determine a phase diagram for the system where four
  different states (free flow, disorder, lanes, gridlock) can be
  distinguished.  Although the number of lanes formed is fluctuating,
  lanes are characterized by a typical density. It is found that the
  basic floor field model overestimates the tendency towards a
  gridlock compared to experimental bounds.  Therefore an anticipation
  mechanism is introduced which reduces the jamming probability.

\end{abstract}

\maketitle


\section{Introduction}
From a physical point of view, human crowds are many-body systems
which show many interesting collective effects. Examples are jamming
at bottlenecks, density waves, flow oscillations, pattern- and lane
formation \cite{Schadschneider01}.  The latter terms the effect that
pedestrians in a crowd segregate spontaneously according to their
desired walking direction. Related phenomena can not only be observed
in pedestrian dynamics but also in other physical systems. For
instance in granular media, the components assort according to their
size when vibrations are applied to the system \cite{granular1}. The
effect occurs particularly in driven systems \cite{Schmittmann199845},
e.g., in complex plasmas \cite{PhysRevLett.102.085003}, molecular ions
\cite{0295-5075-63-4-616} and colloidal suspensions
\cite{PhysRevE.65.021402,nature_colloidal,rex07}. For the latter also
an order parameter for the detection of lanes was introduced which we
are going to adopt for pedestrian crowds
(cf.~Section~\ref{sec:op_def}).  More on the phase diagram and the
lane transition of colloids can be found in Ref.
\cite{lane_colloidol_hyd,lane_colloidol_cont}.

Nevertheless, the phenomenon is best known in pedestrian counterflow
from everyday experience although experimental or empirical
studies are quite rare \cite{NavinW,empir1,empir3,Suma2012248,empir2}.
The effect can be understood as a self-organization process, where
pedestrians try to minimize the contact with other pedestrians,
especially if they have a different desired walking direction. The
mechanism which leads to segregation into distinct lanes is not fully
understood. A possible explanation could be a
fundamental left-right asymmetry, i.e., people prefer to walk either
on the right or on the left. But note that this is not necessary for
the effect. The symmetry can also break spontaneously if certain
assumptions are made, for instance that people tend to follow each
other.

Computer simulations of bidirectional pedestrian movement
(counterflow) are made frequently
\cite{cam10,cam5,cam6,cam8,burstedde01,jam_trans_1,jam_trans_3,jam_trans_2,
jam_trans_4,jam_trans_5,springerlink:10.1007/978-3-540-79992-4_75,Weng2007668,
empir2,socfor1,helbing01,Jiang09,makro_lanes}.  Lane
formation can be reproduced qualitatively
\cite{schadschneider_collective,helbing01} and is often used as a
validation of a model, but there are no quantitative descriptions of
the formation process or the properties of lanes. Even a description
which goes beyond the statement that lanes are present in the system
or even comparisons with empirical data are extremely rare
\cite{empir2,crawling}. Another focus related to bidirectional flow
is on the jamming transition
\cite{jam_trans_1,jam_trans_3,jam_trans_2,jam_trans_4,jam_trans_5},
i.e.\ the transition to a state where no movement is possible
anymore (gridlock). However, there is no empirical evidence for
such a transition and recent experiments showed that the lower
density limit for the transition is larger than 3.5~Persons/m$^2$
\cite{zhang02}.


\section{Definitions}
\subsection{Floor Field Cellular Automaton Model}

\begin{figure}[htbp]
  \centering
  \includegraphics[width=0.25\textwidth]{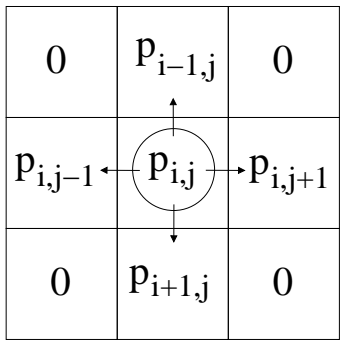}
  \caption{Illustration of the transition probabilities in the von
    Neumann neighborhood for a particle located at $(i,j)$. In the
    Moore neighborhood the four corner cells would have a transition
    probability different from zero.\label{fig:trans_prob}}
\end{figure}

The Floor Field Cellular Automaton (FFCA) model
\cite{burstedde01,kirchner02,KirchnerNS03} is defined on a
two-dimensional square lattice where each cell can be occupied by at
most one particle (pedestrian). In every timestep, each particle is
allowed to stay at its current position or to move to one of the
neighboring cells given that the destination cell is not occupied by
another particle.  This is done synchronously for all particles
(parallel update). The number of nearest neighbors on the lattice can
either be four (von Neumann neighborhood, cf.
Fig.~\ref{fig:trans_prob}) or eight (Moore neighborhood). One can
estimate the area assigned to each cell by taking the reciprocal of
the maximal possible density $\rho_\text{max}$ in a pedestrian crowd.
Although the empirical value of $\rho_\text{max}$ is not known very
well \cite{Schadschneider01}, a common value often used in cellular
automata is $\rho_\text{max}=6.25~\text{Persons}/\text{m}^2$ which
leads to a size of $0.4 \times 0.4~\text{m}^2$ per cell
\cite{kirik07,hua10,dijkstra02}. The length of one timestep can be
estimated by comparing it with the typical reaction time of a
pedestrian. This leads to a value of roughly 0.3 seconds for one
timestep.
The movement of particles is defined by transition probabilities
$p_{ij}$ which (for the model variant \cite{Suma2012248} studied here)
are given by
\begin{align}
  p_{ij} =& \frac{q_{ij}}{\sum_{i,j} q_{ij}}&
  \label{eqn:NormalTransProb}
  \\
  q_{ij} =& \exp\left( k_\mathrm{S} S_{ij} + k_\mathrm{D} D_{ij} -
    k_\mathrm{A} A_{ij} \right) \xi_{ij}
  \label{eqn:StdTransProb}
\end{align}
where $k_\text{S}$ and $k_\text{D}$ are coupling constants. The factor
$\xi_{ij}\in\{0, 1\}$ ensures that movement only takes place at
allowed sites, i.e., $\xi_{ij}=0$ if the cell $(i,j)$ is not
accessible and $\xi_{ij}=1$ otherwise. Inaccessible cells are
wall-cells, cells which are occupied by other pedestrians and cells
which are not in the geometry or the corresponding neighborhood of the
origin-cell. The matrices $S$ and $D$ represent the so-called ``floor
fields'', where $S$ is the static floor field (SFF), $D$ is the
dynamic floor field (DFF) and $A$ is the anticipation floor field
(AFF). The latter is the main difference to the standard FFCA 
\cite{burstedde01,kirchner02} which corresponds to the choice $k_A=0$.
The expression 
\begin{equation}
V_{ij} = -\left(k_\mathrm{S} S_{ij} + k_\mathrm{D}
D_{ij} - k_\mathrm{A} A_{ij}\right)
\end{equation}
in the exponential function of
eqn.~(\ref{eqn:StdTransProb}) can be interpreted as a 
time-dependent potential, i.e., movement is prefered into the
direction of larger $V_{ij}$.

Bidirectional flow is modelled by two different species of
 particles which have opposite preferred walking directions. One
type of particle, which is called ``type~\textit{A}'', is directed
towards the right while ``type \textit{B}'' particles are directed to
the left. In a colloidal suspension these types would correspond to
oppositely charged particles.

The underlying geometry is a rectangle of $W \times L$ cells
where $W$ and $L$ correspond to the width and the length of the
system. A subset of $L$ neighboring cells along the walking direction
is called a row and $W$ neighboring cells perpendicular to the walking
direction are called a column. The particle positions can be described
by a $L \times W$-matrix $\tau_{ij}$ which can be defined as
\begin{equation}
  \tau_{ij}(t) = \begin{cases} 
    +1 &        \text{if~there~is~a~particle~of~type~}A\\
    -1 &        \text{if~there~is~a~particle~of~type~}B\\
    0   &       \text{if~there~is~no~particle}
  \end{cases}
  \label{eqn:def_occupation}
\end{equation}
on the cell $(i,j)$.  The absolute value $|\tau_{ij}|$ corresponds to the
usual binary occupation number.

\subsubsection{Static Floor Field\label{sec:ssf}}
The Static Floor Field (SFF) does not change in time and is not
influenced by the presence of particles. It contains the information
about the preferred direction of motion. Typically it describes the
shortest distance to a destination using some metric. In a counterflow
situation one can give a very simple expression for the SFF in
equation (\ref{eqn:StdTransProb}) which is
\begin{equation}
  S^A_{ij} = \begin{cases}
    +1 & \text{if the cell } (i,j) \text{ is to the \textit{right}}\\
    -1 & \text{if the cell } (i,j) \text{ is to the \textit{left}}\\
    0 & \text{otherwise}
  \end{cases}
\end{equation}
for particles of type $A$ and $S^B_{ij} = - S^A_{ij}$ for particles of
type $B$. This means that different types of particles are affected by
different SFFs.

\subsubsection{Dynamic Floor Field\label{sec:dff}}

The Dynamic Floor Field (DFF) was inspired by the motion of ants, who
leave a pheromone trace \cite{schad_ants}. Other ants are able to
smell this trail and follow it. This concept is adopted in this model:
Each particle which moves from one cell to another leaves a
virtual trace, i.e., the DFF value $D_{ij}$ of the origin cell $(i,j)$
increases by 1. This trace acts attracting on other particles due to
larger transition probabilities according to equation
(\ref{eqn:StdTransProb}). The effect is that particles have a tendency
to follow each other. An important detail is to avoid particles being
attracted by their own virtual trace.  Therefore, in the calculation
of the transition probability in equation (\ref{eqn:StdTransProb}) the
DFF is temporarily reduced by 1 on the last visited cell.  In the same
manner as the SFF one should use two different DFFs $D_{ij}^A$ and
$D_{ij}^B$ which interact only with particles of type $A$ and $B$,
respectively.

The DFF has its own dynamics, namely diffusion and decay. After each
timestep, it is updated according to
\begin{align}
  D_{ij}(t+1) = (1-\delta) \Big[D_{ij}(t) + \frac{\alpha}{4} \Delta
  D_{ij}(t)\Big]\label{eqn:DiffDecCont}
\end{align}
where
\begin{align}
  \Delta D_{ij}(t) &= D_{i,j+1}(t)+D_{i,j-1}(t) \nonumber\\ &+
  D_{i+1,j}(t)+D_{i-1,j}(t)-4D_{i,j}(t)
\end{align}
is a discretization of the Laplace operator.  Equation
(\ref{eqn:DiffDecCont}) can also be obtained by discretizing
\begin{equation}
  \frac{\partial}{\partial t} D_{xy}(t)
  =
  \beta \nabla^2 D_{xy}(t) - \delta D_{xy}(t)
\end{equation}
which is the diffusion equation with diffusion constant $\beta =
\alpha(1-\delta)/4$ and an extra term for the decay. Here $\nabla^2
= \partial^2/\partial x^2 +
\partial^2/\partial y^2$ is the usual Laplace operator in two
dimensions.

\subsubsection{Anticipation Floor Field\label{sec:aff}}

When pedestrians are in a counterflow they usually try to avoid
collisions by estimating the prospective route of pedestrians with
opposite walking direction. This behavior can be imitated by
introducing the anticipation floor field \cite{Suma2012248}. For
particles of type $A$ it is defined as
\begin{equation}
  A_{ij}^A = \sum_{j'} \delta_{1,\tau_{ij'}} \lambda^{d_A(j,j')}
  \label{eqn:antic_def}
\end{equation}
where $\delta_{ij}$ is the Kronecker delta and $\lambda \in (0,1)$ a
parameter which controls the range of anticipation.  $d_A(j,j')$ is
the minimal number of cells which have to be passed if a particle of
type $A$ goes from a cell $(i,j')$ to the cell $(i,j)$, only taking
steps towards its desired walking direction. The field $A_{ij}^B$ for
particles of type $B$ is defined analogously. This definition of
$A_{ij}$ causes a large value of $A_{ij}^X$ ($X \in \{A,B\}$) if there
are many particles of type $X$ which are going to tread the cell
$(i,j)$ in the nearby future. Particles of the other type $Y$ should
avoid this cell, i.e., the AFF acts repulsive (hence the minus sign in
eqn.~(\ref{eqn:StdTransProb})). Note also that only type-$A$ particles
are affected by the AFF generated by type-$B$ particles and vice
versa.

\subsubsection{Update Rules}

The update rules which describe the transfer from one timestep to the
next are given as follows:
\begin{description}
\item[1. Computing the AFF] The AFF is computed in the beginning of
  every timestep according to eqn.~(\ref{eqn:antic_def}) for both
  types of particles.
\item[2. Choosing destination cells]Each particle chooses a
  destination cell according to the transition probabilities given in
  eqn.~(\ref{eqn:NormalTransProb}) and (\ref{eqn:StdTransProb}).
\item[3. Solving conflicts] Conflicts are situations, where $n \geq 2$
  particles have chosen the same destination cell. In this case
  one of the particles is chosen at random to move. For simplicity it
  is assumed that each particle is chosen with equal probability $1/n$. 
\item[4. Movement]Each particle moves to its destination cell. If the
  destination cell is different from the origin cell, the DFF value of
  the latter is increased by 1.
\item[5. Diffusion and decay]The values of the DFF are updated
  according to eqn.~(\ref{eqn:DiffDecCont}). If the boundary conditions are not
periodic, it is assumed that the DFF is zero beyond the boundary, i.e.,
  $D_{ij}=0$ if $i\notin \{1,\cdots,W\}$ or $j\notin \{1,\cdots,L\}$.
\end{description}
Note that rule 3 for dealing with conflicts can be generalized by
introducing a friction parameter \cite{burstedde01,kirchner02,KirchnerNS03}
{or by coupling with evolutionary games
\cite{evolutionary_game_friction,evolutionary_game_friction2,
evolutionary_game_friction3}. 
However, here we will only study the simplest version as described above} since
no indication was found that {the way of dealing with conflicts} has a
qualitative effect on lane formation
\cite{nowak_lanes}.

\subsection{Order Parameter for Lane Formation\label{sec:op_def}}
An order parameter which indicates the presence and absence of lanes
has to detect inhomogeneities parallel to the preferred walking
directions of the particles. A possible approach was given by Yamori
\cite{band_index} who introduced a band index which is basically the
ratio of moving pedestrians in lanes to their total number.
Other attempts for lane detection were made by means of the velocity
profile \cite{burstedde01} and cluster analysis \cite{empir1}. Here we
will use the same order parameter which has been already used in
\cite{rex07} to detect lanes in a colloidal suspension. The particles
there are charged and driven by an external field which determines so
to speak the walking direction. They count for each particle the
number $N_\text{L}$ and $N_\text{O}$ of like and oppositely charged
particles whose lateral distances, i.e., the projection of distance
onto a line perpendicular to the field, is below some threshold $z_c$
(they chose $\frac{3}{4}$ of the particles diameter). Then the
quantity $\phi_n=(N_\text{L}-N_\text{O})^2/(N_\text{L}+N_\text{O})^2$
is assigned to each particle. This value is close to zero if there is
a homogeneous mixture of positive and negative particles and it is
equal to one if there is only one type of particle. The global order
parameter is given by the average $\left< \phi_n \right>$ over
particles.

This concept can be adopted very well in the FFCA. {Taking the spatial
discretness of the FFCA into account} we
chose $z_c$ to be smaller than the diameter of one particle. Hence,
the calculation of $\phi_n$ only considers particles in the same row
due to the discreteness of space. More precisely, to each particle $n$
the value
\begin{equation}
  \phi_n 
  = \left( \frac{\sum_{j=1}^L \tau_{i_n j}}{\sum_{j=1}^L
      |\tau_{i_n j}|} \right)^2
  = \left( \frac{N_{i_n}^A - N_{i_n}^B}{N_{i_n}^A + N_{i_n}^B} \right)^2
  \label{eqn:orderparam_sgl}
\end{equation}
is assigned, where $i_n$ is the vertical position of particle $n$ and
$N_i^A$ ($N_i^B$) denotes the number of particles of type $A$ ($B$) at
row $i$. The global order parameter $\Phi$ is then defined as
\begin{equation}
  \Phi = \frac{1}{N} \sum_{n=1}^N \phi_n
  \label{eqn:orderparam}
\end{equation}
where $N$ is the total number of particles.

Note that the value of $\Phi$ is in general larger than zero, even if
all particles are distributed at random. The mean value of $\Phi$ in
that case is denoted by $\Phi_0$. For small densities $\Phi_0$ can be
large although there is no lane structure in the system. This is taken
into account by defining a reduced order parameter $\tilde \Phi$ as
\begin{equation}
  \tilde \Phi = \frac{\Phi - \Phi_0}{1 - \Phi_0}.
  \label{eqn:reduced_op}
\end{equation}
{Note that this order parameter can in principle become negative!}


\section{Determination of States\label{sec:kd_states}}

If not mentioned otherwise, the number of particles of type $A$ and
$B$ is equal, i.e., $\rho_A=\rho_B=\rho/2$. The parameters are given
as follows:
\begin{align}
  k_\text{S} = 2.5, \quad k_\text{D} = 0, \quad k_\text{A}=0,
  \nonumber
  \\
  \quad \alpha = 0.3, \quad \delta = 0.1,\quad
  \lambda=0.8,\label{eqn:StdParameters} \\\nonumber \quad W=10, \quad
  L=100\text{.}
\end{align}
Considering the cell size of 40~cm~$\times$~40~cm, this corresponds to
a corridor of 40~m~$\times$~4~m.  The boundary conditions are periodic
in walking direction and open perpendicular to it, i.e., there is a
virtual wall at row $0$ and $W+1$. We use the von Neumann
neighborhood, i.e., each cell has four nearest neighbors.

\subsection{Definition of Different States\label{sec:def_states}}

\begin{figure}[htb]
  \includegraphics[width=0.49\textwidth]{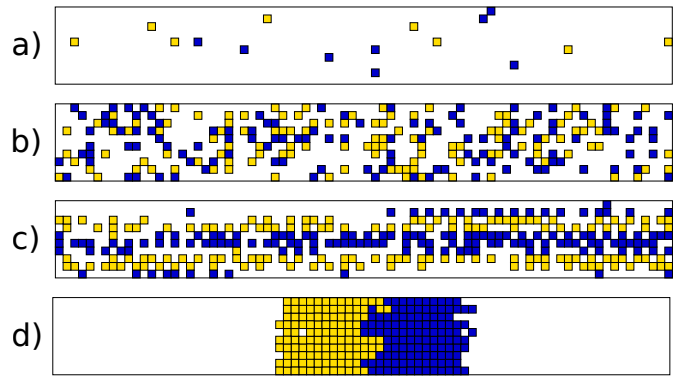}
  \caption{(Color online) Definition of states: a) Free flow, b) Disorder, c)
Lanes, d) Gridlock. \label{fig-states}}
\end{figure}

The patterns observed in the system can be classified into at
least four different states, which are defined as follows (see also
Fig.~\ref{fig-states}):
\begin{description}
\item[Free flow] The particle density is very low such that the
  average distance is too large for a considerable interaction between
particles. This leads to a small reduced order parameter $\tilde \Phi$ and a
large mean velocity.
\item[Disorder] This state is characterized by a large homogeneity of
  particles which leads to a small order parameter.
\item[Lanes] Almost every row of cells contains only one type of
  particle apart from a small number of exceptions caused by
  fluctuations. $\tilde \Phi$ is large in this state.
\item[Gridlock] This denotes a complete jam in the system. Particles
  of type $A$ and $B$ clog each other such that no movement is
  possible in the desired direction. The reduced order parameter is 
usually negative and the velocity $v$ is zero.
\end{description}

\subsection{Simulation Procedure\label{sec:sim_proc}}

The simulations were done as follows: Initially, the $N=\rho W L$
particles are distributed randomly on the lattice. Then the simulation starts
for different densities $\rho \in \{0.05,0.06,0.07,\ldots,1.00\}$ and for
different coupling constants $k_\text{D} \in \{0.0,0.1,0.2,\ldots,7.5\}$. For
each pair $(\rho,k_\text{D})$ of parameters the simulation was
repeated at least 100 times. The simulation runs were stopped if one
of the following conditions is fulfilled:
\begin{enumerate}
\item The flow $J$ averaged over the last 50 timesteps is lower than
  $\frac{1}{2 W L}=5 \times 10^{-4}$. {This criterion indicates
the formation of a gridlock.}
\item The maximum and minimum value $\Phi_\text{max}$ and
  $\Phi_\text{min}$ of the order parameter $\Phi$ in the last 1000
  timesteps fulfills the following condition
\begin{equation}
\frac{\Phi_\text{max}-\Phi_\text{min}}{\Phi_\text{max}+\Phi_\text{min}}
< 0.1\,,
\label{lanecondition}
\end{equation}
{which indicates the formation of lanes}.
\item The number of timesteps exceeds the density dependent value
  $T_\mathrm{max}=20000\sqrt{\rho}$. {This typically happens for disordered or
free flow states.}
\end{enumerate}
If a simulation is stopped due to the second or third
condition, the velocity $v$, flow $J$ and order parameter $\Phi$ are
averaged over the last 1000 timesteps.  Since the first condition
means that the average number of particles which moved towards their
desired direction is less than $\frac{1}{2}$, it indicates the
occurrence of a gridlock. For a given set of parameters, the fraction
of simulation runs which are aborted due to the first condition
defines the jam probability $P_\text{Jam}$. Condition 2 indicates the
occurrence of lanes. However, it does not contain any
restrictions for the absolute value of the order parameter $\Phi$,
because more than 99\% of all values for $\Phi$ are anyway larger than
$0.85$ given that $\Phi$ converges.

Finally, the last condition ensures that the computation time for a
simulation run is finite. The value of $T_\mathrm{max}$ is based on the
experience that lanes and gridlocks are formed much more quickly, if at all.
The factor $\sqrt{\rho}$ takes into account
that the system takes more time to evolve into a stationary state if
the number of particles is increased. Using $\sqrt{\rho}$ instead
of $\rho$ leads to a fast decrease in simulation time at low densities
where the system does evolve neither into a jammed state nor a lane
state.

\subsection{Results}
\subsubsection{Jam Probability\label{sec:jam_prob}}
\begin{figure*}[htbp]
  \centering
  \includegraphics[width=0.49\textwidth]{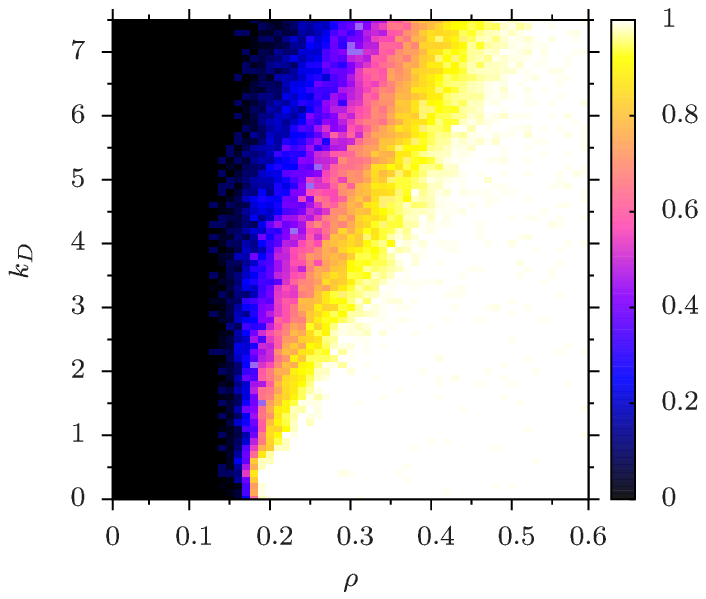}
  \includegraphics[width=0.49\textwidth]{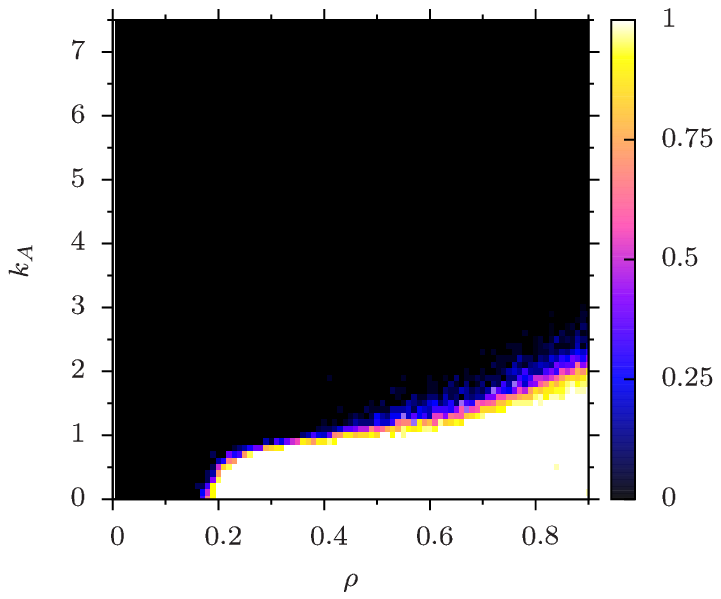}
  \caption{(Color online) Jam probability $P_\mathrm{Jam}$ as function of
    the density $\rho$ and the coupling constants $k_\mathrm{A}$
    and $k_\mathrm{D}$. Left:
    $k_\mathrm{D}>0$, $k_\mathrm{A}=0$.  Right: $k_\mathrm{D}=0$,
    $k_\mathrm{A}>0$. \label{fig:jam_kd_map}}
\end{figure*}
As a first result one observes that in general the jam probability
$P_\text{jam}$ increases both with an increasing density and a
decreasing coupling constant $k_\text{D}$ and $k_\text{A}$,
respectively (Fig.~\ref{fig:jam_kd_map}).

This result is not very surprising since an increasing density means a
larger number of opposing particles which can block each
other. Additionally, there is less space to avoid such collisions. An
increasing $k_\text{D}$ means that it becomes more likely for a
particle to follow another particle of the same kind and thus
collisions of type-$A$ and type-$B$ particles are avoided. That the
AFF prevents collisions is more obvious, because it prevents by
definition particles from coming too close to opposing particles in
the same row.

The simulations clearly indicate that the AFF is more capable to
prevent jams compared to the DFF. If the AFF is turned off ($k_A=0$,
Fig.~\ref{fig:jam_kd_map} left) all simulations with a density larger
than 0.5 evolve into a gridlock. Even for $\rho=0.2$ one has a
non-vanishing jam probability. Anticipation leads to a vanishing jam
probability for a sufficiently large coupling constant $k_\text{A} >
3$ at almost every density.

\subsubsection{Order Parameter\label{sec:kd_op}}

\begin{figure*}[hbtp]
  \centering
  \includegraphics[width=0.49\textwidth]{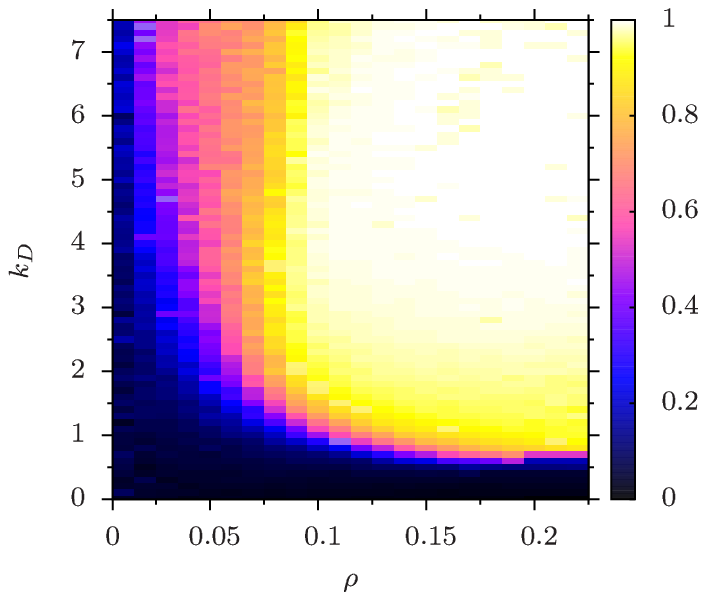}
  \includegraphics[width=0.49\textwidth]{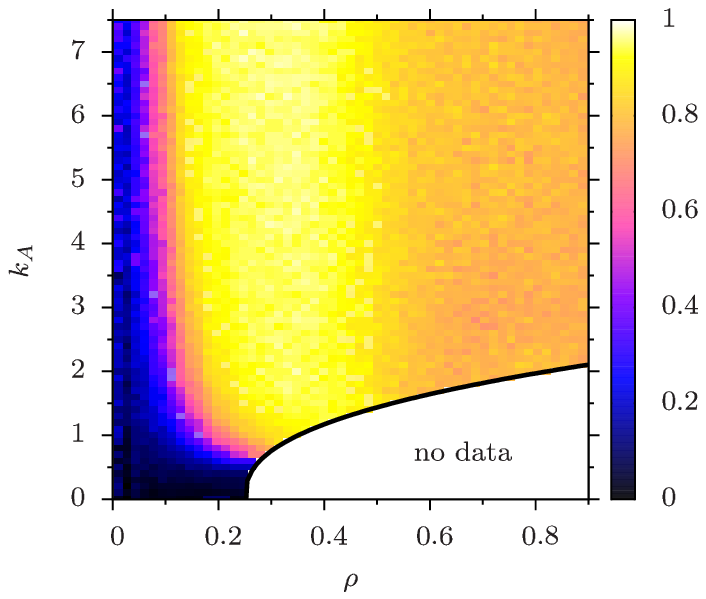}
  \caption{(Color online) Reduced order parameter $\tilde \Phi$ as function
    of the density $\rho$ and the coupling constants. Left:
    $k_\mathrm{D}>0$, $k_\mathrm{A}=0$.  Right: $k_\mathrm{D}=0$,
    $k_\mathrm{A}>0$.\label{fig:op_kd_map}}
\end{figure*}
Now we consider only those simulation runs which are \textit{not}
aborted due to the detection of a gridlock. In general, the averaged
reduced order parameter $\tilde \Phi$ becomes larger for increasing
densities $\rho$ and coupling constant (see Fig.~\ref{fig:op_kd_map}).

As expected, for $k_\text{D}=k_\text{A}=0$ the reduced order parameter
is close to zero for all densities. The fact that it is slightly above
zero can be explained by the increased number of sidesteps a particle
has to perform if it has contact to a particle of different kind.
These collisions happen more frequently if the number of type-$A$ and
type-$B$ particles in a specific row is similar. The resulting
sidesteps lead to a decreasing stability of homogeneous states.

By increasing the coupling constants the order parameter $\tilde \Phi$
shows a specific behavior which depends on the density. The dependence
is qualitatively the same for lanes that are formed by the DFF and
AFF, but for the latter there are more data for larger densities
available due to a smaller jam probability.  One can see that a
certain minimal density is needed to reach an order parameter which is
close to 1. The transition from a disordered state to a state with
almost perfect lanes becomes sharper for increasing density. For
densities $\rho > 0.2$ the transition can only be observed at a few
data points because most of the simulations evolve into a gridlock for
small coupling constants. In particular there is no stable disordered
state for large $\rho$ and the system evolves either in a jammed state
or in a state with almost perfect lanes.

\subsubsection{Dependence on the System Size}

Since the behavior is quite sensitive to the size of the
system, we want to discuss briefly the influence of the dimensions.
\begin{figure}[htbp]
  \includegraphics[width=0.49\textwidth]{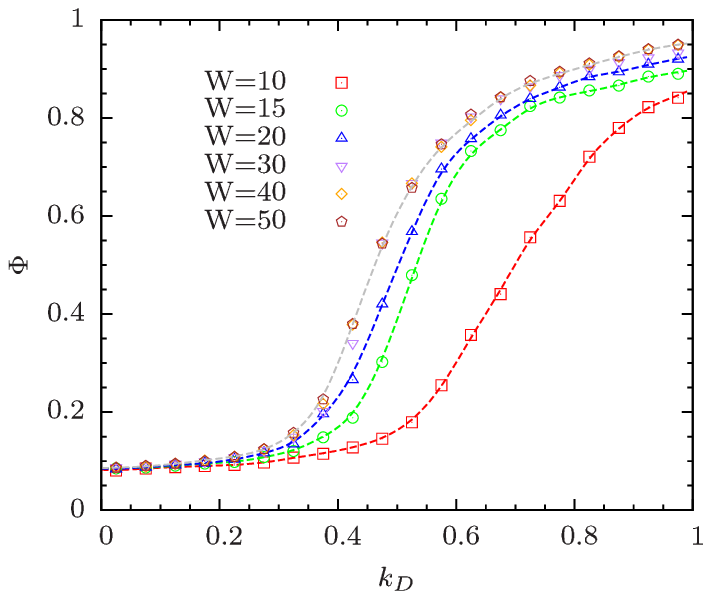}
  \caption{(Color online) Dependence of the order parameter on the
    system size at density $\rho=0.15$.\label{fig:op_size}}
\end{figure}
The results for the order parameter $\Phi$, which can be found in
Fig.~\ref{fig:op_size}, show that $\Phi$ increases with increasing
width $W$ of the corridor if the length $L$ remains constant. This
increase seems to converge at a width of about $W=30$, since the value
of the order parameter does not change if $W$ is increased further.

\begin{figure}[htbp]
  \includegraphics[width=0.49\textwidth]{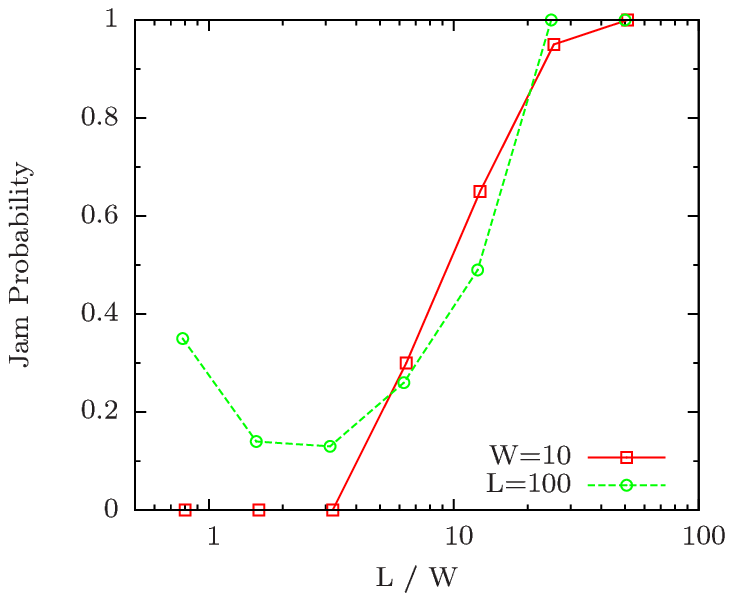}
  \caption{(Color online) Dependence of the jam probability on the
    system size  ($\rho=0.2$, $k_\mathrm{D}=2$).\label{fig:jam_size}}
\end{figure}

A similar scaling of the system length $L$ could not be performed
due to the increasing jam probability in that case 
(cf. Fig.~\ref{fig:jam_size}).
As a rough approximation one can say that $P_\text{jam}$ does only
depend on the ratio $L/W$ and is monotonously increasing with this
ratio. But this is only justified for smaller systems and $L/W >
1$. For large values of $W$ one can observe that $P_\text{jam}$ is 
non-monotonic whereas $P_\text{jam}$ increases with increasing $W$.


\section{Properties of the Lane State}

\subsection{Distribution of Densities}

In this Section we discuss how the particles distribute in the system
after the formation of lanes. The data are taken after a simulation
run was stopped due the {lane condition (\ref{lanecondition})} as
described in Section~\ref{sec:sim_proc}.

\begin{figure}[htb]
  \includegraphics[width=0.49\textwidth]{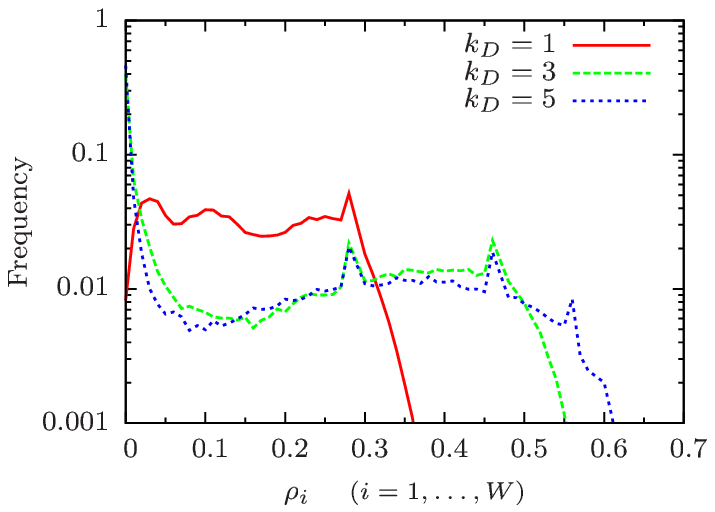}
  \caption{(Color online) Distribution of densities in the different
    rows. The global density is $\rho=0.15$, the bin size is
    0.01.\label{fig:dens_distr_kd}}
\end{figure}
Figure~\ref{fig:dens_distr_kd} shows the distribution of densities
$\rho_i$ in the different rows $i \in \{1,\ldots,W\}$. If the
particles were equally distributed over the system, one would expect a
sharp peak in the distribution at the position of the global density
$\rho$. This is for instance the case in a system without DFF, even if
anticipation is included. But since the DFF implies an attractive
interaction between particles, the number of rows which are not
occupied as well as the number of rows with larger density grows. This
can be observed in the distribution as a maximum at $\rho_i = 0$
and a larger support for increasing coupling constant $k_\text{D}$.
The latter means that the mobility of particles and thus the average
velocity is reduced. This will be analyzed in more detail in
Section~\ref{sec:velocities}.

Another feature of the distribution is the existence of characteristic lane
densities which do neither depend on the coupling
constant nor on the global density. This surprising result was tested for
different coupling constants $k_\text{D} \in [1,7]$ and for different
(global) densities $\rho \in [0.1, 0.3]$. These lane densities become visible
as peaks in the distribution at $\rho_1^* = 0.28$,
$\rho_2^* = 0.46$ and $\rho_3^* = 0.56$. The different peaks correspond to
states with different number of lanes, i.e., $\rho_j^*$ corresponds to a
configuration with $j+1$ lanes. Narrow lanes have a higher density than wide
lanes. Note that the tendency of forming more lanes ($N_\mathrm{lanes}>2$)
increases with increasing coupling constant $k_D$. Therefore, for small
coupling constants (e.g., $k_\mathrm{D}=1$) only 2-lane configurations occur
resulting in a single peak whereas for large coupling constants (e.g.,
$k_\mathrm{D}=5$) also 3- and 4-lane configurations occur resulting in three
peaks.

Furthermore, {under certain circumstances there are} additional peaks
in the distribution. {At small densities and large coupling constants
$k_\text{D}$ almost all particles of equal type are located in the same row.
This causes additional density dependent peaks. For example at $\rho=0.1$ and
a system of size $100 \times 10$ there are 50 particles of each
type and the maximal possible density in one row of the lane state is
$\rho_i=0.5$. Therefore, if $k_\text{D}$ ist large, there is a peak at
$\rho^*=0.5$ in the distribution. A similar effect can be found for large
global densities. Then there can be an additional maximum beyond $\rho_3^*$
which can even have a larger value.}

\subsection{Velocities\label{sec:velocities}}
Since lane formation is a way of separating different types of
particles which leads to less opposing traffic one would expect that
lanes increase the mean velocity $v$ in the system, i.e., that
$v(k_\mathrm{A})$ is monotonously increasing. However, this is only
true for lanes which are formed due to the AFF, i.e., $v(k_\mathrm{A})$ is
monotonously increasing \cite{nowak_lanes}.

\begin{figure}[htbp]
  \includegraphics[width=0.49\textwidth]{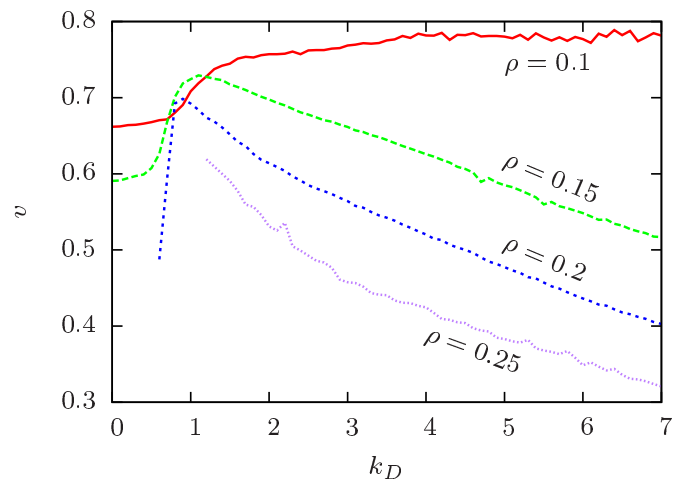}
  \caption{(Color online) Velocities in dependence of the coupling
    constant $k_\text{D}$. The data are gained by the simulation
    procedure described in
    Section~\ref{sec:sim_proc}.\label{fig:velocity}}
\end{figure}

As one can see in Fig.~\ref{fig:velocity}, the dependence of $v$ on
$k_\text{D}$ is not always monotonous.
The DFF acts attracting on particles of the same type and thus the
local densities in a row can be large, as seen in the previous section
(cf. Fig.~\ref{fig:dens_distr_kd}). This means in general a decrease
of velocity, because particles are hindered by particles of the same
kind. Thus there is a decrease in $v(k_\text{D})$. Note that a state with
maximal velocity has not necessarily a maximal order parameter.

Note that not for all densities a maximum is present. For instance at
$\rho=0.1$ the global density does not suffice to create a local
density which is larger than 0.5. Hence, each particle has on average
at least one empty cell in front and the velocity does not
decrease. If the density is too large one has also no maximum because
there are no data available for small $k_D$ due to a large jam
probability, i.e., the curve $v(k_\text{D})$ starts at some
$k_\mathrm{D}>0$ and can be monotonously decreasing in this case.

\subsection{Stability and Lifetime\label{sec:lifetime}}
In agreement with the observations in empirical studies
  \cite{zhang02} the general structure of lanes in the
simulations fluctuates in time.
  The number of lanes changes due the vanishing of lanes or merging of
  neighbouring lanes to a broader one. In this section we will
analyze this phenomenon. Previously, simulations were initialized 
in a disordered state and evolved into a lane state. 
Now we take the inverse approach, i.e., the 
simulations are started in a state where lanes already have formed and
we wait until this structure is destroyed. The lifetime of lanes will
serve as a measurement for their stability. Since the stability
depends on the lane width we use different initial conditions with 2,
3 and 4 lanes. The simulation starts with one of these initial conditions 
and is stopped if either the number of timesteps exceeds 
$T_\text{max}=1.5 \times 10^5$ or if the order parameter $\Phi$ falls
below $\frac{1}{\sqrt{2}}$. The first condition is necessary to
perform the simulation runs in an acceptable time. This might distort
the result for states which have lifetimes of order $T_\text{max}$.
However,  $T_\text{max}=1.5 \times 10^5$ corresponds to more than 
10 hours in real time since one timestep corresponds to about $0.25$ 
seconds.

\begin{figure}[htpb]
  \includegraphics[width=0.49\textwidth]{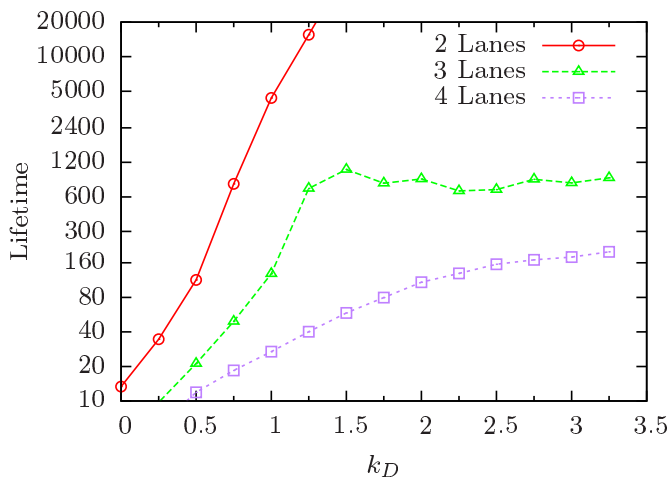}
  \caption{(Color online) Lifetime of lanes for different initial
    conditions for a density of $\rho=0.25$.\label{fig:lifetime}}
\end{figure}
The results are shown in Fig.~\ref{fig:lifetime}. The two-lane
lifetime increases roughly exponentially with increasing
$k_\text{D}$. It is not surprising that the stability is decreased if
the number of lanes is increased because then the fraction of
particles which are at the interface between two lanes becomes
larger. The three-lane lifetime only increases up to a value of
$k_\text{D} \approx 1.25$ where one can see a kink in the curve.  For
larger values of $k_\text{D}$, the lifetime stays almost constant at a
value of about 1000 timesteps. Finally, the four-lane configuration
has a lifetime which smoothly converges to a value of about 250.

Simulation with anticipation but without the DFF show a different
behavior. The influence of the number of lanes on the lifetime is
smaller. The two-lane configuration has still the largest lifetimes,
but the difference in the stability of the three- and four-lane
configuration becomes very small for increasing coupling
constants. In all three cases the lifetime grows faster than
exponential with $k_A$.

The fate of a system which has lost its stability depends on the
density and the value of the coupling constant
(cf. Fig.~\ref{fig:jam_kd_map} and \ref{fig:op_kd_map} from
Section~\ref{sec:kd_states}). For large densities and a small coupling
constant, the system evolves into a jam state. For small densities and
a small coupling constant, the system evolves into a disordered
state. If the coupling constant is large, i.e., at least large enough
to prevent jams, the system can form lanes again. If the initial
configuration consists of more than two lanes one can sometimes
observe that the lanes are merging. In that case the final
configuration consists of two lanes. Usually this happens only in
systems with large coupling to the DFF, because then the attraction
between two lanes with the same walking direction is strong.  In most
cases the order parameter $\Phi$ falls below $\frac{1}{\sqrt{2}}$,
such that the simulation is aborted and the correct lifetime is
taken. Anyway, sometimes it happens that lanes are merging without an
appreciable decrease of $\Phi$ and thus the condition for stability is
useless. The frequency of these incidents increases with increasing $k_D$ and
make up to 50\%.
Fortunately, one can easily identify
these cases in the data,
because the lifetime of a 2-lane configuration is much larger and thus
their is a huge gap (at least one magnitude) in the distribution of
lifetimes. Hence, we only take into account the results below this gap.

\vspace{5mm}


\section{Open Boundary Conditions}
\subsection{Two Groups Passing Each Other\label{sec:groups_kd}}
\begin{figure}[htbp]
  \includegraphics[width=0.49\textwidth]{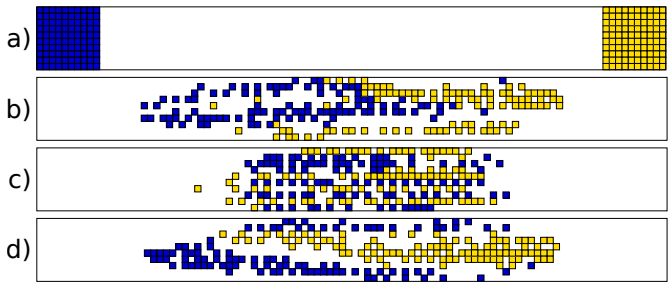}
\caption{(Color online) Two groups
walking past each other.
a) Initial conditions. b)-d) Snapshot of the simulation at timestep 75.
b) $k_D=2, k_A=0$,
c) $k_D=0, k_A=2$,
d) $k_D=k_A=2$.
\label{fig:groups_svg}
}
\end{figure}

For this scenario we start the simulation with two groups of particles
at the very left and the very right end of the system (cf.
Fig.~\ref{fig:groups_svg}a).
Each group contains 100 individuals of type $A$ and $B$, respectively.
They walk in their desired walking direction until they reach the
boundary of the system. There they are absorbed and removed from the
system.  The simulation ends when the flow becomes zero, i.e., if
either all particles are removed or a gridlock is formed.  The
simulation was repeated 5000 times for each pair of parameters
$(k_\text{D}, k_\text{A}) \in \{0,1,2,3,4\}^2$.

The first question is whether the groups are able to pass each other
or not. If both dynamic floor field and anticipation is turned off,
the system always evolves in a gridlock state. If $k_D$ becomes
larger the jam probability is much reduced, e.g., for $k_D=2$ (but
$k_A=0$) about 50\% of simulation runs form a gridlock. For $k_D \ge
0$ and $k_A \ge 2$ no gridlocks were observed at all. As before we
only take into account those simulation runs which did not
{end in a gridlock.}

In order to quantify the tendency of collisions, we introduce a
collision index $n_c$ as follows: A ``collision'' is defined as a pair
of two neighboring cells, where the left one is occupied by a type-$A$
particle and the right one by a type-$B$ particle. Let $N_\text{Col}$
be the number of those pairs in the system. Then the collision index
is defined as
\begin{align}
  n_c = 2 \frac{N_\text{Col}}{N}
  \label{eqn:n_c}
\end{align}
where $N$ is the total number of particles.

\begin{figure}[htbp]
  \includegraphics[width=0.49\textwidth]{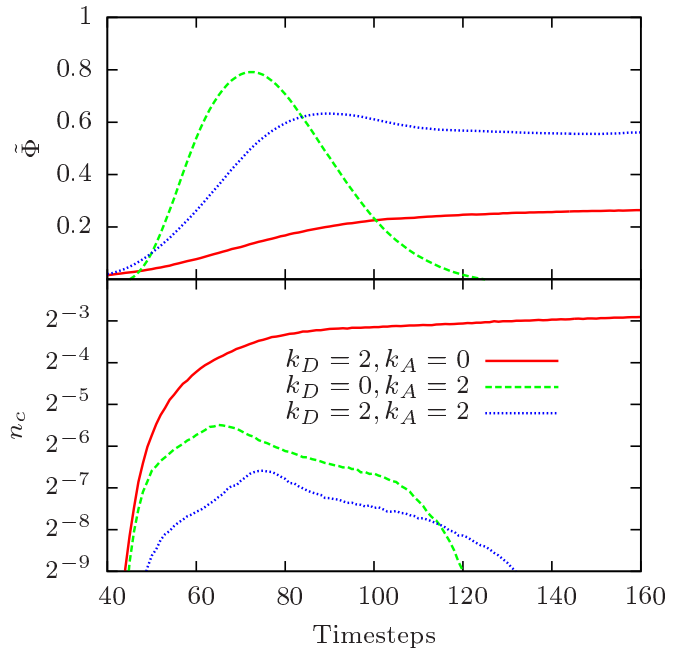}
  \caption{(Color online) Order parameter $\Phi$ and collision index
    $n_c$ from eqn.~(\ref{eqn:n_c}) versus time for two groups walking
    past each other.\label{fig:groups_op_collisions}}
\end{figure}

One can observe that the dynamics shows qualitative differences
depending on the choice of the coupling constants. Without
anticipation it is strongly influenced by collisions between opposing
particles (cf.\ Fig.~\ref{fig:groups_svg}b
and~\ref{fig:groups_op_collisions}).  Even after most of the particles
have already left the system there are almost always small groups of
particles which form a jam that dissipates after longer time.
Therefore, the collision index $n_c$ fades very slowly. Lanes can only
be formed after both groups get in direct contact with each other, but
they are less distinctive compared to those in the periodic system.
This leads to a quite small order parameter.
With anticipation collisions are much less likely, but there are also
difference in the dynamics depending on whether the DFF is turned
on. If not, many thin lanes are formed during the contact
(Fig.~\ref{fig:groups_svg}c). After the groups have passed each other
these lanes are destroyed immediately. This can also be observed in
terms of the order parameter in
Fig.~\ref{fig:groups_op_collisions}. Probably the most realistic results
were attained by a combination of AFF and DFF. The tendency of collisions
is reduced further. Lanes can be seen clearly even if the order
parameter is a bit reduced compared to the system with AFF only. This
can be explained by lanes which are often not parallel to the walking
direction (Fig.~\ref{fig:groups_svg}d). Their number is much decreased
and one usually observes 2 or 3 lanes. The lane structure is kept
after the groups have passed each other.

\subsection{Random Insertion of Particles}
Now the particles are inserted and removed dynamically
with a certain insertion probability $\beta_\text{in}=\beta$ on each
cell next to the boundaries. At the left (right) boundary only
type-$A$ (type-$B$) particles are inserted. Like in the previous
section, the removal of particles will always happen, i.e., with
probability $\beta_\text{out}=1$ at the left and right boundary,
respectively. The left boundary absorbs only type-$B$ particles
whereas the right boundary absorbs only type-$A$ particles.
\begin{figure}[htbp]
  \vspace{3mm}
  \includegraphics[width=0.49\textwidth]{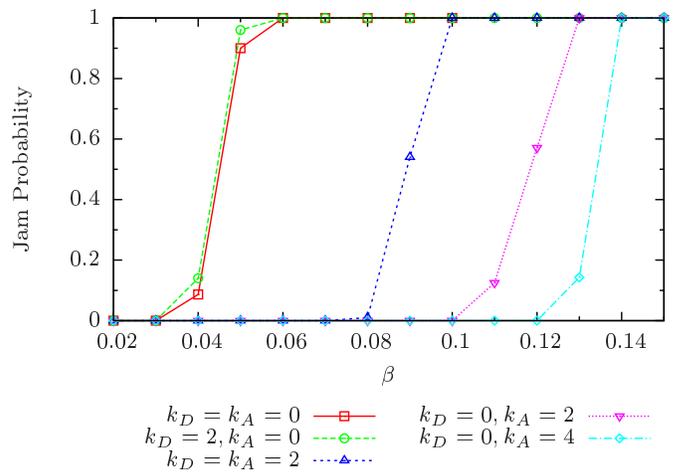}
  \caption{(Color online) Jam probability at timestep 20000 for open boundary
conditions.\label{fig:obc_jam}}
\end{figure}

Fig.~\ref{fig:obc_jam} shows the probability that the system evolves
into a gridlock before timestep 20000. It turns out that the avoidance
of jams is much harder as in the periodic system. The DFF does not
help very much to prevent jams but astonishingly it rather supports
the formation of a jam.

Anticipation helps much to avoid gridlocks and even for an insertion
probability of $\beta = 0.12$ gridlocks can be avoided. We find a
linear dependence between $\beta$ and $\rho$. The constant of
proportionality depends slightly on the exact value of the coupling
constants and ranges from 2.5 to 2.8. This means that the maximal
density in the system without forming a gridlock is about $\rho =
0.34$.

Before gridlocks are formed, the system usually forms lanes for
sufficiently large coupling constants. The order parameter is lower
than in the periodic case, but comparable to the findings in the
previous section (Fig.~\ref{fig:groups_op_collisions}).

{It is not surprising that systems with open boundary conditions
  show a slightly different gridlock behaviour from that with periodic
  boundaries. In the periodic case the flow in a system near a
  gridlock becomes very small. In contrast, in the open system still
additional particles are fed into the system. Due to particles moving
up "from behind" the chances for the gridlock to resolve are 
strongly reduced.}


\section{Conclusions}

We have studied the phenomenon of lane formation in a cellular
automaton model for pedestrian dynamics. In contrast to previous studies,
we have obtained for the first time quantitative results. 
This was possible by introducing an order parameter which has
been adapted from the analysis of colloidal suspensions.
It has been shown that this order parameter is suitable for the
detection of lanes, at least for the simple scenario of a straight
corridor considered here. 

It is found that at least four different states can be distinguished,
namely free flow, disordered flow, lane formation and gridlock.
Using the order parameter we have mapped out the parameter range
of the model for which lane formation is possible.
Also other quantities like density and velocity show a very 
characteristic behavior in these scenarios.

Most models for pedestrian dynamics show a strong tendency towards
gridlock which appear for large densities. This is rather unrealistic
and was not observed in experimental studies. A model extension
which includes an anticipation mechanism was found to show a more
realistic behavior by suppressing the formation of gridlocks, at
least in the case of periodic boundary conditions.

{Although we have focussed on the floor field model and its
variants we want to emphasized that the techniques developed here
can also be applied to most other models.}


\section*{Acknowledgements}
This work was supported by the project {\tt Hermes} funded by the
Federal Ministry of Education and Research (BMBF) Program on
''Research for Civil Security - Protecting and Saving Human Life''
under grant no.\ 13N9960 and the Bonn-Cologne Graduate School of
Physics and Astronomy (BCGS).



%

\end{document}